\documentclass[8.5pt,twoside,twocolumn]{article}
\usepackage[super,sort&compress,comma]{natbib}
\usepackage{mhchem}
\usepackage{times,mathptmx}
\usepackage{nicefrac}
\usepackage{units}
\usepackage{ulem}
\usepackage{numprint}
\usepackage[squaren, Gray, cdot]{SIunits}
\usepackage{sectsty}
\usepackage{balance}

\usepackage{graphicx} 
\usepackage{lastpage}
\usepackage[format=plain,justification=raggedright,singlelinecheck=false,font=small,labelfont=bf,labelsep=space]{caption}
\usepackage{fancyhdr}
\begin{document}

\fancypagestyle{plain}{
\renewcommand{\headrulewidth}{1pt}}
\renewcommand{\thefootnote}{\fnsymbol{footnote}}
\renewcommand\footnoterule{\vspace*{1pt}%
\hrule width 3.4in height 0.4pt \vspace*{5pt}}
\setcounter{secnumdepth}{5}

\makeatletter
\def\subsubsection{\@startsection{subsubsection}{3}{10pt}{-1.25ex plus -1ex minus -.1ex}{0ex plus 0ex}{\normalsize\bf}}
\def\paragraph{\@startsection{paragraph}{4}{10pt}{-1.25ex plus -1ex minus -.1ex}{0ex plus 0ex}{\normalsize\textit}}
\renewcommand\@biblabel[1]{#1}
\renewcommand\@makefntext[1]%
{\noindent\makebox[0pt][r]{\@thefnmark\,}#1}
\makeatother
\renewcommand{\figurename}{\small{Fig.}~}
\sectionfont{\large}
\subsectionfont{\normalsize}

\fancyfoot{}
\fancyhead{}
\renewcommand{\headrulewidth}{1pt}
\renewcommand{\footrulewidth}{1pt}
\setlength{\arrayrulewidth}{1pt}
\setlength{\columnsep}{6.5mm}
\setlength\bibsep{1pt}

\twocolumn[
  \begin{@twocolumnfalse}
\noindent\LARGE{\textbf{Cluster of red blood cells in microcapillary flow: hydrodynamic versus macromolecule induced interaction}}
\vspace{0.6cm}

\noindent\large{\textbf{Viviana  Claver\'{i}a,\textit{$^{a,b}$} Othmane Aouane,\textit{$^{a,c,d,e}$} Marine Thi\'ebaud,\textit{$^{c,d}$} Manouk Abkarian,\textit{$^{b}$} Gwennou Coupier,\textit{$^{c,d}$} Chaouqi Misbah,\textit{$^{c,d}$} Thomas John,\textit{$^{a}$} and Christian Wagner\textit{$^{a}$}}}\vspace{0.5cm}

\noindent \normalsize{We present experiments on RBCs that flow through micro-capillaries under physiological conditions. The strong flow-shape coupling of these deformable objects leads to a rich variety of cluster formation. We show that the RBC clusters form as a subtle imbrication between hydrodynamics interaction and adhesion forces because of plasma proteins, mimicked by the polymer dextran. Clusters form along the capillaries and macromolecule-induced adhesion contribute to their stability. However, at high yet physiological flow velocities, shear stresses overcome part of the adhesion forces, and cluster stabilization due to hydrodynamics becomes stronger. For the case of pure hydrodynamic interaction, cell-to-cell distances have a pronounced bimodal distribution. Our 2D-numerical simulations on vesicles captures the transition between adhesive and non-adhesive clusters at different flow velocities.}
\vspace{0.5cm}
 \end{@twocolumnfalse}
  ]

\section{Introduction}

\footnotetext{\textit{$^{a}$~Experimental Physics, Saarland University, 66123, Saarbr\"{u}cken, Germany. Tel: +49 (0)681 302 - 3003; E-mail: c.wagner@mx.uni-saarland.de}}
\footnotetext{\textit{$^{b}$~Centre de Biochimie Structurale, CNRS UMR5048, INSERM U1054, Universit\'{e} de Montpellier, 29 rue de Navacelles, 34090 Montpellier, France}}
\footnotetext{\textit{$^{c}$~Universit\'{e} Grenoble Alpes, LIPHY, F-38000 Grenoble, France. }}
\footnotetext{\textit{$^{d}$~CNRS, LIPHY, F-38000 Grenoble, France. }}
\footnotetext{\textit{$^{e}$~LMPHE, URAC 12, Facult\'{e} des Sciences, Universit\'{e} Mohammed V- Agdal, Rabat, Morocco.}}

\begin{figure}[h!]
	\includegraphics[width = 1 \linewidth]{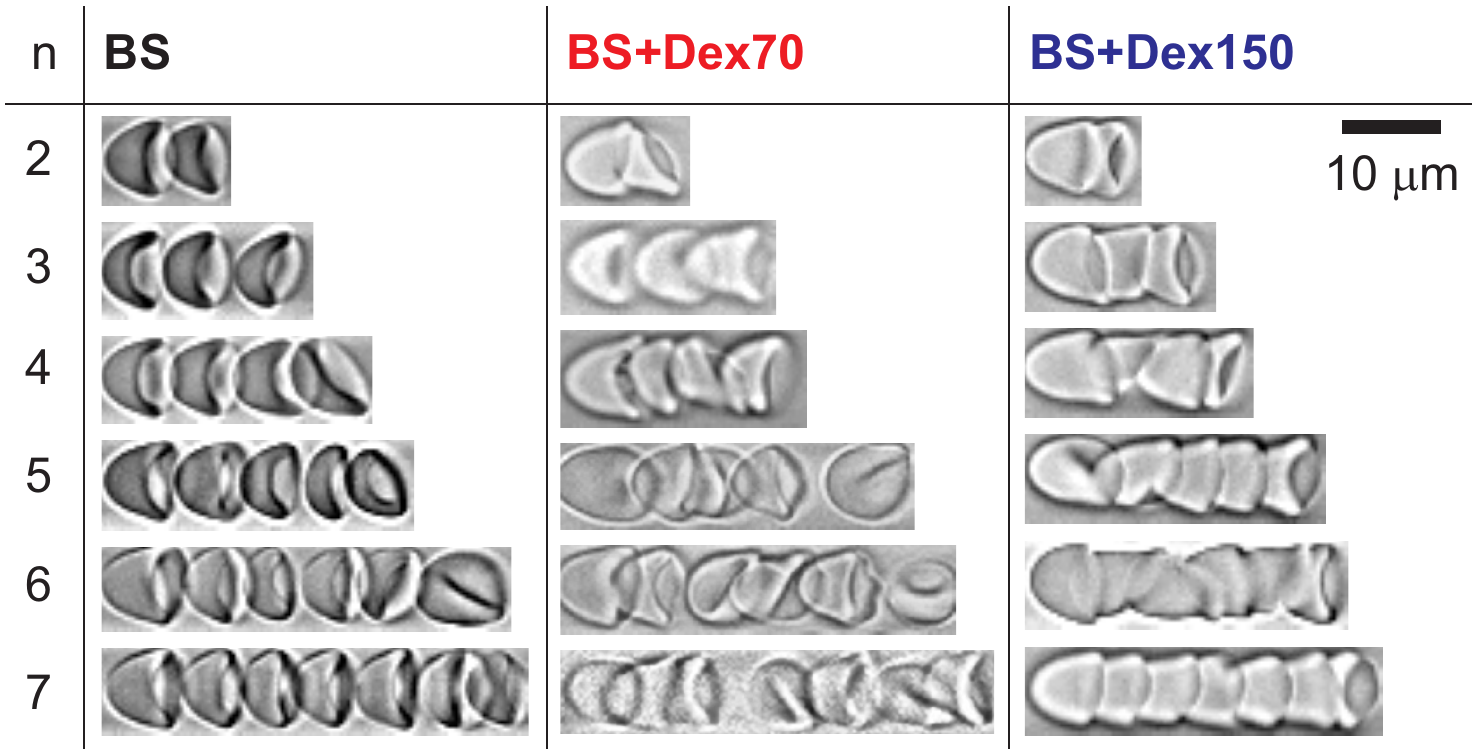}
	\caption{\label{fig:geo_clust} Snapshot of the typical red blood cell (RBC) cluster configurations of type $n$, which contain $n$ cells for an imposed $\Delta P =$ \unit{100}{mbar}. In the physiological base solution (BS), the RBCs are separate from one another and mainly in the axis-symmetric parachute or asymmetric slipper-like shape. In the base solution with the macromolecule dextran 150 (BS+Dex150), the cells adhere closely to one another and are mainly in a bullet shape. RBCs in BS+Dex70 are in an intermediate configuration.}
\end{figure}

The flow of blood in our vascular system is determined by the behaviour of its main constituents: the red blood cells (RBCs). Physically, they can be modelled as soft deformable objects such as vesicles, and their shapes strongly couple to the flow. Even if the typical volume concentration of RBCs in the micro-capillary network can be as low as one percent locally  \cite{klitzman1979microvascular}, observations show that well-organized clusters tend to form (cp. Fig.\,\ref{fig:geo_clust}). The physical origin of this cluster formation can be either the long-ranged hydrodynamic interaction \cite{gaehtgens1979motion,tomaiuolo2012red,schmid1980interaction,mcwhirter2009flow,ghigliotti2012and} or a short-range aggregation mechanism, which is caused by the plasma macromolecules \cite{Brust2014}. The latter relates to the so-called \textit{rouleaux} formation \cite{baskurt2011red}. RBCs in plasma at rest always form clusters that look like stacks of coins, i.e. rouleaux. This reversible aggregation mechanism was explained using models based on either the bridging of the macromolecules between the cells, or the depletion effect \cite{Neu2002,Steffen2013}. One unanswered question is: how big is the relative contribution of the hydrodynamic interaction compared to the macromolecule-induced interaction on the cluster formation in a confined flow? Naturally, any additional aggregation mechanism can be a micro-circulatory risk factor and indicates diseases such as inflammation, ischemic strokes or pathological thrombus formation \cite{di2001prognostic,rampling2004influence,steffen2011stimulation}.

For hard-sphere colloidal systems, it has been recently shown that both bridging and depletion can simultaneously affect the process of crystallization and self-assembly \cite{feng2015re,jee2015colloidal,fantoni2015bridging}. However, there is some quantitative disagreement between numerical predictions and experiments on the kinetics of phase transitions in colloidal systems, and it is predicted that hydrodynamic interactions must be considered to resolve this discrepancy \cite{raduEPL2014}. There is some knowledge on the mechanisms of hydrodynamic interaction in a confined flow of hard spheres \cite{cuiprl2004} and spherical droplets suspensions \cite{beatus2007anomalous} where even phonon-like excitations can be observed, but much less is known in the case of soft deformable objects. However, there are some recent and notably detailed numerical predictions that soft objects should have a much stronger tendency to form clusters and obey a richer dynamic than hard systems \cite{mcwhirter2011deformation}. Indeed, red blood cells can be considered as model objects that are provided by biology to study the flow of soft objects in confinement, where the physical parameters of the objects, such as deformability \cite{abkarian2015importance} or chemical interaction potentials, are as important as the hydrodynamics around them. A recent review on the behaviour of blood in micro-capillaries in vitro was given by Guido et al.\cite{guido2009microconfined}. 

At rest, RBCs are biconcave discs with a diameter of  \unit{7.5 \hspace{.1cm} to \hspace{.1cm} 8.7}{\micro\meter} and \unit{1.7 \hspace{.1cm} to \hspace{.1cm} 2.2}{\micro\meter} in thickness \cite{diez2010shape} encapsulating a solution with a typical viscosity of $\eta_{in} = $\unit{5}{mPas} at \unit{37}{\celsius}. Its membrane is composed of a self-assembled fluid lipid bilayer with a thickness of approximately \unit{4}{\nano\meter} and an attached network of proteins called cytoeskeleton. Local and global area changes on the membrane are small, making it possible to treat the membrane as a two-dimensional incompressible material\cite{skalak1973strain}. Additionally, RBC membrane elasticity is attributed to the inner-attached protein network and then two independent modes of deformation are necessary to describe its elastic state: the simple shear at constant surface area and the out-of-plane bending. Other important factors that determine RBC deformability include the viscosity of the red blood cell membrane, viscosity contrast between the cytoplasm of the cell and their suspending media, as well as the RBC volume to surface area ratio, where the high deformability is promoted by the extensive cell area excess (larger area compared to the area of sphere of the same volume). In this way, RBCs can squeeze through capillaries with diameters as small as \unit{2}{\micro\meter} \cite{shelby2003microfluidic}.

The rouleaux formation of RBCs in plasma is mostly caused by the plasma protein fibrinogen. The typical adhesion energy $\epsilon$ at healthy concentrations of fibrinogen is around \unit{5}{\micro\joule\per\meter\squared} \cite{Brust2014}. In this particular study, we will mimic this condition in terms of adhesion energy and we will include an additional case with a higher adhesion energy using two dextran solutions with different molecular weights, based on data reported by Steffen et al.\cite{Steffen2013}. Of course, even if dextran allows to mimic the typical adhesion energies, there are still many subtle differences between the effect of dextran and fibrinogen. Aggregation and disaggregation might have a different dynamic for different polymers or proteins \cite{Lee2016} and it is also known that dextran adsorbs on the RBC surface, changing its elasticity to some extend \cite{Rampling1972}. We use ${\unit{20}{\milli\gram\per\milli\litre}}$ of dextran 70 (molecular weight of \unit{70}{kDa}) to induce an interaction energy of $\epsilon =$\unit{4.8}{\micro\joule\per\meter\squared} and $\unit{20}{\milli\gram\per\milli\litre}$ of dextran 150 (molecular weight of \unit{150}{kDa}) to induce an interaction energy of $\epsilon =$\unit{12}{\micro\joule\per\meter\squared} \cite{Steffen2013}. Additionally, the interplay between different flow velocities and consequently different capillary numbers, with and without macromolecules inducing aggregation in the suspending media is also investigated.

In former studies on cluster formation, either no macromolecules were added and only hydrodynamic effects were present \cite{ghigliotti2012and}, or the effects of the macromolecules were tested in channels that were smaller than the RBC diameter with the effect of hydrodynamic attraction minimized \cite{Brust2014}. In this study, we allow both mechanisms to take effect and we observe that both are relevant, but they form clusters of notably different geometrical forms. From Fig. \ref{fig:geo_clust} one can deduce already a definition of a pure hydrodynamic cluster or a polymer induced, adhering one. In the BS solutions, cells always have some finite surface to surface distance whilst in the case of dextran most cells stick closely to each other.  Our two-dimensional numerical simulations predict that the flow velocity significantly affects the cluster configuration in the macromolecule case.

\section{Materials and methods}

\subsection{Experimental}

Blood samples were collected via finger prick from two healthy donors. To remove all substances except RBCs, the sample was centrifuged at 3000 rpm for 3 minutes at room temperature, and the liquid phase and buffy coat were removed by aspiration. Then, the sample was re-diluted in \unit{1}{ml} of physiological buffer solution (PBS, phosphate buffered saline, Invitrogen, Darmstadt, Germany) including \unit{1}{mg/ml} of bovine serum albumin (BSA, Polysciences, Warrington, USA) to eliminate crenation and preserve the biconcave shape \cite{ponder1971hemolysis}. The BSA in PBS is called the base solution (BS). To prepare the solution with the macromolecules that mimic healthy and pathological levels of fibrinogen, \unit{20}{mg} of dextran 70 or 150 (Sigma Aldrich, Taufkirchen, Germany) were diluted in \unit{1}{ml} of BS. These solutions are called BS+Dex70 and BS+Dex150, respectively. A SU-8 micro-channels master mold was used to build the PDMS-based microfluidic device. It comprises 25 parallel channels that start and end at an inlet reservoir of approximately \unit{1}{\milli\meter} in height. The channels have a rectangular cross-section of \unit{11.9 \pm 0.3}{\micro\meter} $\times$ \unit{9.7 \pm 0.3}{\micro\meter} in width and height, respectively, and all channels have a length of $\sim$\unit{4.0}{\centi\meter}. A high-precision pressure device (Elveflow, OB1, France) was connected to an external 1.5 ml reservoir that contained the RBC sample. Through a polyethylene tube, this reservoir is connected to the inlet of the microfluidic device to inject the cells into the channels at three constant pressure drops  $\Delta P = $ \unit{(50, 100, 1000) \pm 0.5 \%}{mbar} and one additional pressure drop of \unit{20 \pm 0.5 \%}{mbar} for the BS. For rectangular channels, in the absence of cells, this would lead to a maximum flow velocity ${v}_{max}$ given by \cite{bruus2008theoretical}

\begin{equation}
{v}_{max} = \frac{4h^2\Delta P}{\pi^{3} \eta_{out} L} \sum_{n,odd}^{\infty} \frac{1}{n^3}\left[1-\frac{1}{\cosh\left(\frac{n \pi w}{2h} \right)} \right]\sin\left({\frac{n\pi}{2}}\right)
\label{max_vel}
\end{equation}

where $h$, $w$ and $L$ are the channel height, width and length, respectively, and $\eta_{out}$ is the suspending media viscosity.
The maximum flow velocities in our channels for BS, BS+Dex70 and BS+Dex150 solutions are calculated using Eq. (\ref{max_vel}), with $\eta_{out} = 1$,  $1,67$ and $\unit{1,80}{\milli\pascal\second}$ for BS, BS+Dex70 and BS+Dex150 solutions respectively. These results are summarized in Table\,\ref{tab:table0}. Please remark that the viscosity of the dextran solutions without RBCs did not show any Non-Newtonian, i.e. shear thinning behaviour in our commercial cone plate rheometer for shear rates between 10 and 1000 $s^{-1}$. 
	
		\begin{table}[h]
			\small
			\caption{\ Maximum flow velocity per solution calculate from Eq. (\ref{max_vel}) for the pressure drops used in this experiment in our rectangular cross-section channels.}
			\label{tab:table0}
			\begin{tabular*}{0.5\textwidth}{@{\extracolsep{\fill}}llll}
				\hline
				$\Delta$P &  ${v}_{max, BS}$ &   ${v}_{max, BS+Dex70}$ &  ${v}_{max, BS+Dex150}$ \\
				($\textrm{\milli}$bar)&($\textrm{\milli\meter\per\second}$)&$(\textrm{\milli\meter\per\second}$)&$(\textrm{\milli\meter\per\second}$)\\
				\hline
				20 & $0.41$ & $-$ &  $-$  \\
				50 & $1.03$ & $0.62$ & $0.57$ \\
				100 & $2.07$ & $1.24$ & $1.15$ \\
				1000 & $20.65$ & $12.37$ & $11.47$ \\
				\hline
			\end{tabular*}
		\end{table}
		
A better quantity to describe the relative flow strength than by the mean flow velocities only is the capillary number, as we will see below. It compares the flow forces  with the bending resistance: 

\begin{equation}
		Ca  =  \frac{\eta_{out} \bar{v} a^{2}}{k_{B}} 
	\label{ca}
\end{equation}

where $\bar{v}$ is the average RBC velocity, and $a = 3$ $\micro\meter$ is a characteristic length of the system, in our case the cell radius.

Images of the flowing cells were taken at 30 frames per second for the lower values of the pressure drop and at 300 frames per second for a $\Delta P =$  \unit{1000}{mbar} using a high-magnification oil immersion objective (60\textsf{x}, NA = 1.4). With the help of a motorized x–y stage, four different positions were observed along the channels. Each position cover a field of view of \unit{0.25}{\milli\meter} in width and is located at \unit{0}{\milli\meter}, \unit{1}{\milli\meter}, \unit{2}{\milli\meter}, and \unit{10}{\milli\meter} from the entrance of the channels that we will now call positions 0, 1, 2 and 10, respectively. A home-made image-processing routine enabled the detection of individual cells. We define two cells to be a part of a cluster if their center-to-center distance is smaller than or equal to 1.5 times the length of one single cell as proposed by Tomaiuolo \textit{et al.} \cite{tomaiuolo2012red}. However, this value is heuristically selected and not based on theoretical considerations. Furthermore, as we always consider the length of the actual cell which becomes more and more elongated with the flow strength, this definition of the length of a cluster changes with the flow strength. However, our statistical analysis below with the measured probability density functions of cluster length distribution gives a posteriori justification of this simple definition.

\subsection{Numerical}

\subsubsection{Membrane model ~~} Vesicles are closed phospholipid bilayer membranes encapsulating an inner fluid and are suspended in outer fluid with the corresponding dynamic viscosities $\eta_{in}$ and $\eta_{out}$. Their membrane exhibits three kinds of strain: i) stretching; ii) tilting and iii) bending. In general, the curvature is the only strain governing the shape of non-spherical vesicles \cite{Helfrich1973}. The total elastic bending energy can therefore be written in the most simplified form as the addition of two contributions, the mean and the Gaussian curvatures
\begin{equation}
	E_c = \frac{\kappa}{2} \oint{\left(c_1 + c_2 -c_0\right)^2 dA} + \kappa_G \oint{c_1 c_2 dA}
	\label{eq::Helfrich-tmp1}
\end{equation}
where $c_1 = 1/R_1$ and $c_2=1/R_2$ are the principal curvatures, $c_0$ is the spontaneous curvature, $dA$ is the area element, $\kappa$ and $\kappa_G$ are the bending and the Gaussian curvature moduli. In our two-dimensional model, the Gaussian curvature is an irrelevant constant owing to the property $\oint c ds = \oint \frac{d\theta}{ds} ds = 2\pi$, where $ds$ is the arc length and $d\theta$ is the corresponding tangential angle. By considering only symmetrical membranes \textendash~same composition of phospholipids~\textendash \ the spontaneous curvature term can be disregarded (i.e. $c_0 = 0$). The incompressibility and in-extensibility of the membrane are fulfilled by the use of a Lagrange Multiplier $\zeta$. %
The subsequent total energy is expressed as
\begin{equation}
E_c = \frac{\kappa}{2} \oint{c^2 ds} + \oint{\zeta ds}
\label{eq::Helfrich}
\end{equation}
The two-dimensional membrane force is then obtained by calculating the functional derivative of eq. \ref{eq::Helfrich}:
\begin{equation}
\mathbf{f}_{struct \mapsto fluid} =  - \frac{\delta{E_c}}{\delta \mathbf{X}} =\kappa \left[\frac{\partial^2{c}}{\partial s^2} + \frac{c^3}{2}\right]\mathbf{n}  - c \zeta \mathbf{n} +
\frac{\partial{\zeta}}{\partial{s}}\mathbf{t}
\label{eq::membrane_force}
\end{equation}
where $\mathbf{n}$ and $\mathbf{t}$ are namely the normal and the tangent unit vectors at a position vector $\mathbf{X}$ belonging to the membrane.
\subsubsection{Cell-Cell aggregation model ~~} The RBCs suspended in a plasma solution are able to aggregate and form three-dimensional complex structures called "rouleaux" assimilable to a stack of coins. In non-pathological conditions, the rouleaux formation is reversible under the effect of disaggregating forces, such as shearing forces (e.g. due to an external flow), electrostatic repulsion akin to the negative charge of the red blood cell membrane and membrane strain (e.g. bending). %
In the early fifties, Asakura and Oosawa introduced the depletion theory to explain the interaction force between two large spherical particles due to smaller particles (macromolecules) flowing around \cite{Asakura1958}. The exclusion of macromolecules from the area between the large spheres induces an osmotic pressure acting as an attractive force leading to aggregation. The second model is the bridging theory, which assumes that the RBC membrane adsorbs surrounding macromolecules and forms bridges with the neighbouring RBCs, leading to the formation of an aggregation if the adhesion forces exceed that of the disaggregating forces\cite{Chien1975,Maeda1986}.
The origin of the mechanism of aggregation is still a puzzling problem, although some evidences seem to be favoring the depletion theory. Neu and Meiselmann provided a theoretical model based on the depletion theory that includes the effect of the polymer concentration, the polymer physiochemical properties on depletion layer thickness and the polymer penetration depth into the RBC glycocalyx\cite{Neu2002}. The theoretical predictions were in good agreements with the few available experimental data points on the quantification of RBCs adhesion energy obtained via micro-pipette aspiration technique \cite{Buxbaum1982}. Recently, a single cell force spectroscopy approach has been used to measure the adhesion energy between two RBCs in the presence of macromolecules Dextran \cite{Steffen2013}, confirming the robustness of the depletion model proposed by Neu and Meiselmann. To mimic the effect of RBCs aggregation on viscoelastic properties of blood flow, Liu and Liu used a Morse potential to simplify the expression of the total cell-cell interaction energy \cite{Liu2006} resulting in an important reduction of the number of control parameters. An alternative approach based on a Lennard-Jones potential
\begin{equation}
\label{eq::LJpotential}
\phi(r) = 4\epsilon\left[\left(\frac{\sigma}{r}\right)^{12}-\left(\frac{\sigma}{r}\right)^{6}\right]
\end{equation}
was previously used by our group to simulate the role of macromolecules on the flow of RBCs in micro-circulation \cite{Brust2014}. %
The weak depletion attraction and the strong electrostatic repulsive forces at large and short distances are the negative derivative of the intercellular interaction potential
\begin{equation}
\label{eq::LJforce}
\mathbf{f}^{\phi}(\mathbf{X}) = -\int_{\sum_{j \neq i}{\partial \Omega_{j}}} {\frac{\partial{\phi(r)}}{\partial{r}}\frac{\mathbf{r}}{r}ds(\mathbf{Y})}
\end{equation}
where $\epsilon$ and $\sigma$ denote the surface energy and the zero force length distance.  $\mathbf{r} = \mathbf{X} - \mathbf{Y}$, $r = \|\mathbf{X} - \mathbf{Y}\|$, and
$\mathbf{X}$ and $\mathbf{Y}$ are two position vectors belonging to the $i-th$ and $j-th$ membrane ($\partial \Omega$), respectively. The well-depth is the equivalent of the surface energy and is deduced directly from the single cell force spectroscopy measurements \cite{Steffen2013,Brust2014}. In order to reproduce a cell surface to cell surface distance comparable to the 25 nm that have been reported in the literature for adhering RBCs at rest in Dex70 solution \cite{chien1973}, but to keep the effort due to finite numerical resolution limited, we chose a zero force distance $r = \sigma * 2^{1/6} =$ 480 nm and which is still below the optical resolution in the experiments.
\subsubsection{Boundary integral method~~}
In the absence of vesicles, the motion of the fluid in circular channels is dictated by the Poiseuille's flow profile formula
\begin{equation}
\begin{array}{rcl}
{v_x}^\infty & = & v_{max}\left[1-\left(\frac{y}{W/2}\right)^2\right] \\
{v_y}^\infty & = & 0 \\
\end{array}
\end{equation}
where $\mathbf{v}_{max}$ is the mid-plane velocity of the Poiseuille flow and $w$ is the channel width. The presence of deformable passive interfaces (i.e. vesicles) will disturb the base Poiseuille flow and as result, the motion equation needs to be solved. The Stokes equations are reformulated in a boundary integral representation \cite{pozrikidis1992boundary} and coupled with the vesicle's membranes. The specific Green's functions satisfying the no-slip boundary at the two boundaries are calculated using the images method and a Fourier transform \cite{Liron1976,Marine2013}. The velocity along the membrane is expressed as
\begin{multline}
\label{eq::velocity_integral_depletion}
\mathbf{v}(\mathbf{X_0}) = \frac{2}{1+ \lambda}\mathbf{v}^\infty(\mathbf{X_0}) + \\
\frac{1}{2 \pi \eta_{out} (1+ \lambda)} \int\limits_{\sum_{i}{\partial \Omega_{i}}}{\mathbf{G}(\mathbf{X},\mathbf{X_0})[\mathbf{f}_{struct \mapsto fluid}(\mathbf{X}) + \mathbf{f}^{\phi}(\mathbf{X}) ] ds(\mathbf{X})} \\%
+ \frac{(1-\lambda)}{2 \pi (1+ \lambda)}\int\limits_{\sum_{i}{\partial \Omega_{i}}}\mathbf{v}(\mathbf{X}) \cdot \mathbf{T}(\mathbf{X},\mathbf{X_0}) \cdot \mathbf{n}(\mathbf{X})ds(\mathbf{X}) 
\end{multline}
where $\mathbf{v}^\infty$ is the velocity of the imposed Poiseuille flow in the absence of the vesicles, $\mathbf{G}$ and $\mathbf{T}$ are namely the Green's function for two parallel walls and its associated stress tensor, and $\lambda$ is the viscosity contrast between the inner and outer fluids, namely $\eta_{in}$ and $\eta_{out}$. The position of the nodes is then advected in time using an Eulerian explicit scheme. This model has been developed and used by our group to study the flow of single cells and clusters in the micro-circulation \cite{Othmane2014,Brust2014}, and to investigate the rheological properties of blood \cite{Marine2014}.

\section{Results and discussions}
\subsection{Overview}
The average flow velocity of the red blood cells for each pressure drop of each solution are presented in Table\,\ref{tab:table1}. From Eq. (\ref{ca}) the associated capillary number  $Ca$ for each average velocity was calculated (Table\,\ref{tab:tableca}).

Due to rouleaux formation and sedimentation of RBCs at the inlet reservoir in macromolecule induced adhesion solutions, only RBCs in BS solution were possible to flow at 20 mbar.

	\begin{table}[h]
		\small
		\caption{\ Average RBCs velocity per pressure drop and suspending media obtained by image analysis for each solution.}
		\label{tab:table1}
		\begin{tabular*}{0.5\textwidth}{@{\extracolsep{\fill}}llll}
			\hline
			$\Delta$P & $\bar{v}_{BS}$ &  $\bar{v}_{BS+Dex70}$ &  $\bar{v}_{BS+Dex150}$ \\
			($\textrm{\milli}$bar)&($\textrm{\milli\meter\per\second}$)&($\textrm{\milli\meter\per\second}$)&($\textrm{\milli\meter\per\second}$)\\
			\hline
			20  & $0.25$ $\pm$ $0.01$ & $-$ &  $-$  \\
			50  & $0.73$ $\pm$ $0.01$ & $0.45$ $\pm$ $0.01$ & $0.35$ $\pm$ $0.02$ \\
			100 & $1.39$ $\pm$ $0.08$ & $0.92$ $\pm$ $0.07$ & $0.76$ $\pm$ $0.01$  \\
			1000 & $15.75$ $\pm$ $0.50$ & $10.47$ $\pm$ $0.40$ &$8.75$ $\pm$ $0.42$  \\
			\hline
		\end{tabular*}
	\end{table}

\begin{table}[h]
	\small
	\caption{\ Capillary number $Ca$ for each pressure drop and solution.}
	\label{tab:tableca}
	\begin{tabular*}{0.5\textwidth}{@{\extracolsep{\fill}}llll}
		\hline
		
		 & $Ca_{BS}$ &  $Ca_{BS+Dex70}$ &  $Ca_{BS+Dex150}$ \\
		\hline
		
		20  & $5.6$ $\pm$ $0.2$ & $-$ &  $-$  \\
		50  & $16.4$ $\pm$ $0.2$ & $16.9$ $\pm$ $0.4$ & $14.2$ $\pm$ $0.8$ \\
		100 & $31.3$ $\pm$ $1.8$ & $34.6$ $\pm$ $2.6$ & $30.8$ $\pm$ $0.4$  \\
		1000 & $354.4$ $\pm$ $11.3$ & $393.4$ $\pm$ $15.0$ &$354.4$ $\pm$ $17.0$  \\
		\hline
	\end{tabular*}
\end{table}
The velocity of the cells are inversely proportional to the medium viscosity. As a consequence, the capillary numbers are similar for all solutions for each pressure drop, implying a valid comparison among the different solutions for each pressure drop. 

Fig.\ref{fig:geo_clust} shows an overview of different experimentally observed clusters type $n$ for $\Delta P = \unit{100}{mbar}$, where \textit{n} is the number of cells that belongs to the referred cluster. The shape of a RBC in a cluster formed in the BS solution is typically an axis-symmetric parachute or a slipper and the cells within a cluster are clearly separated from one another. This result is different for the RBC in the BS+Dex70 solution, where most of the cells adhere closely to their neighbours and their parachute shape is deformed. This result is even more pronounced in the case of BS+Dex150 where the cells have a bullet shape. There are no macromolecules in the BS solution, therefore, we conclude that the clusters are formed by hydrodynamic interactions only, whereas the cells in the solutions with dextran adhere closely because of the macromolecules. For the cluster $n = 7$ in the BS+Dex70 solution (cp. Fig.\ref{fig:geo_clust}), as an example, it becomes apparent that this cluster may consist of two adhering cells followed by three other adhering ones. This result indicates that this cluster is a hydrodynamic cluster consisting of two macromolecule-induced clusters of 2 and 3 cells each.

\subsection{Span wise positions}
\begin{figure}
	\centering
	\includegraphics[width = 1 \linewidth]{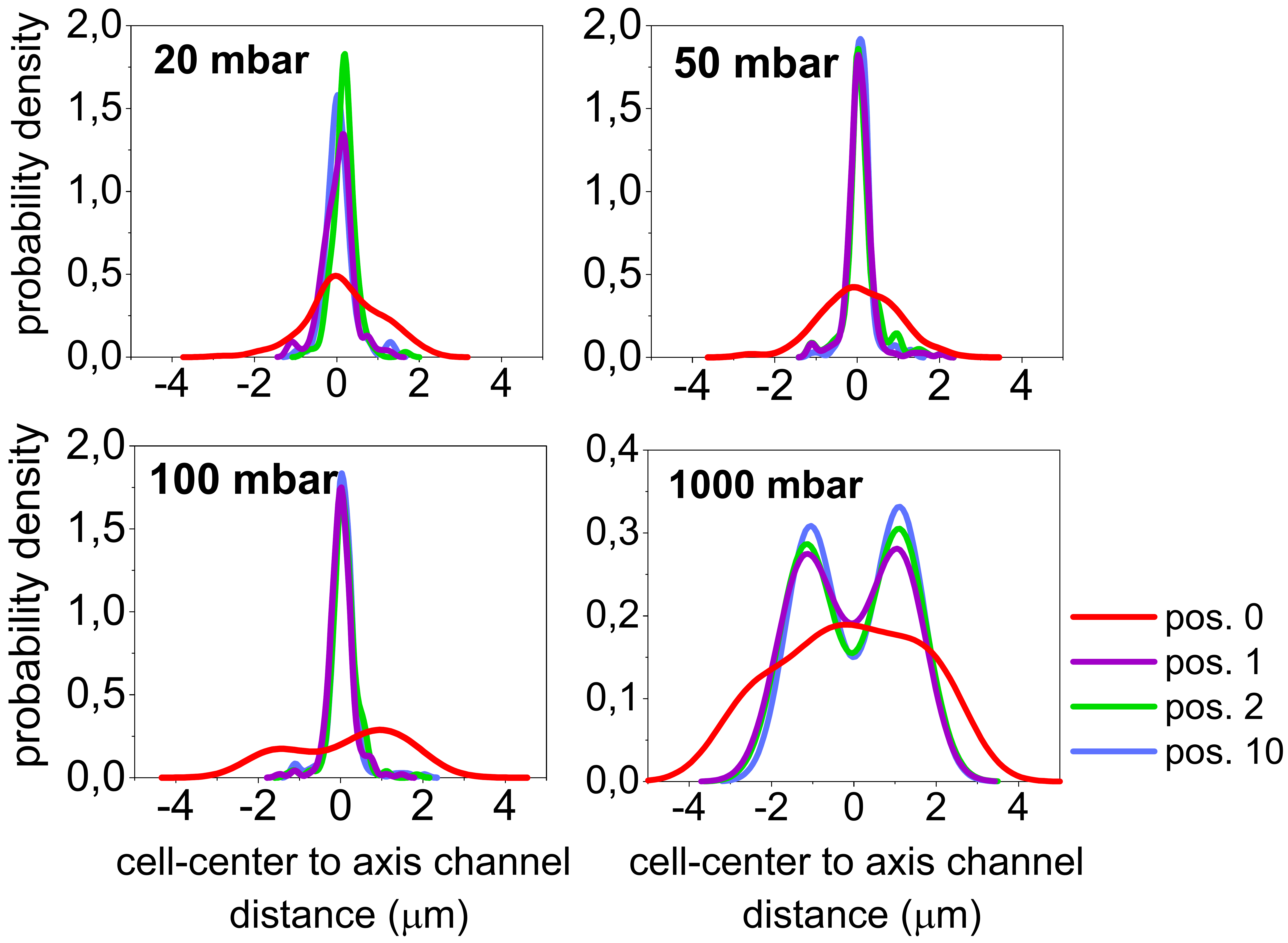}	
	\caption{\label{fig:dist}[Color online]  Probability density of the cell centre to channel axis distance for $\Delta P$ = 20, 50, 100 and \unit 1000 {mbar} in BS.}
\end{figure}

Before we quantify the clustering process in detail, we first characterize the span wise position of the cells. At the entrance, the cells enter the channels with a distance from their centre to the channel axis in a random distribution, independent of the pressure drop and the solution. Some of the cells enter closer to the walls while others closer to the centre. In Fig.\ref{fig:dist}, the probability density of the cell-center to channel axis distance is plotted for the cell flow in BS solution. A centred axial alignment is observed for $\Delta P = 20$ to $\unit{100} {mbar}$ from position 1 to 10. A different arrangement is observed for $\Delta P = \unit{1000} {mbar}$, where the cells tend to have a non-centred alignment with the axis of the channel located symmetrically up and above the channel axis at a distance of $1.45 \pm 0.35\,\textrm{\micro\meter}$ from position 1, and remain almost invariant until position 10. A similar behaviour is observed for the macromolecule cases. Thus, the behavior of the cells is apparently steady from position 1 onwards and position 0 has only been taken into account to quantify and compare the relative occurrence of the cluster formation between the different positions while cluster length, shape as well as the discrimination between hydrodynamic and macromolecule-induced clusters have been quantified from position 1 to 10.  
It is also worth noting that the span wise cell-positioning is apparently related to the shape of the cell that depends on the flow velocity. At smaller velocities, the cells are mainly in the axial-centered parachute configuration, but at higher velocities we found them to be in an off-centred slipper configuration predominantly.

\subsection{Cluster formation along the channels}

\begin{figure}[h!]
 	\centering
 	\includegraphics[width = 0.8 \linewidth]{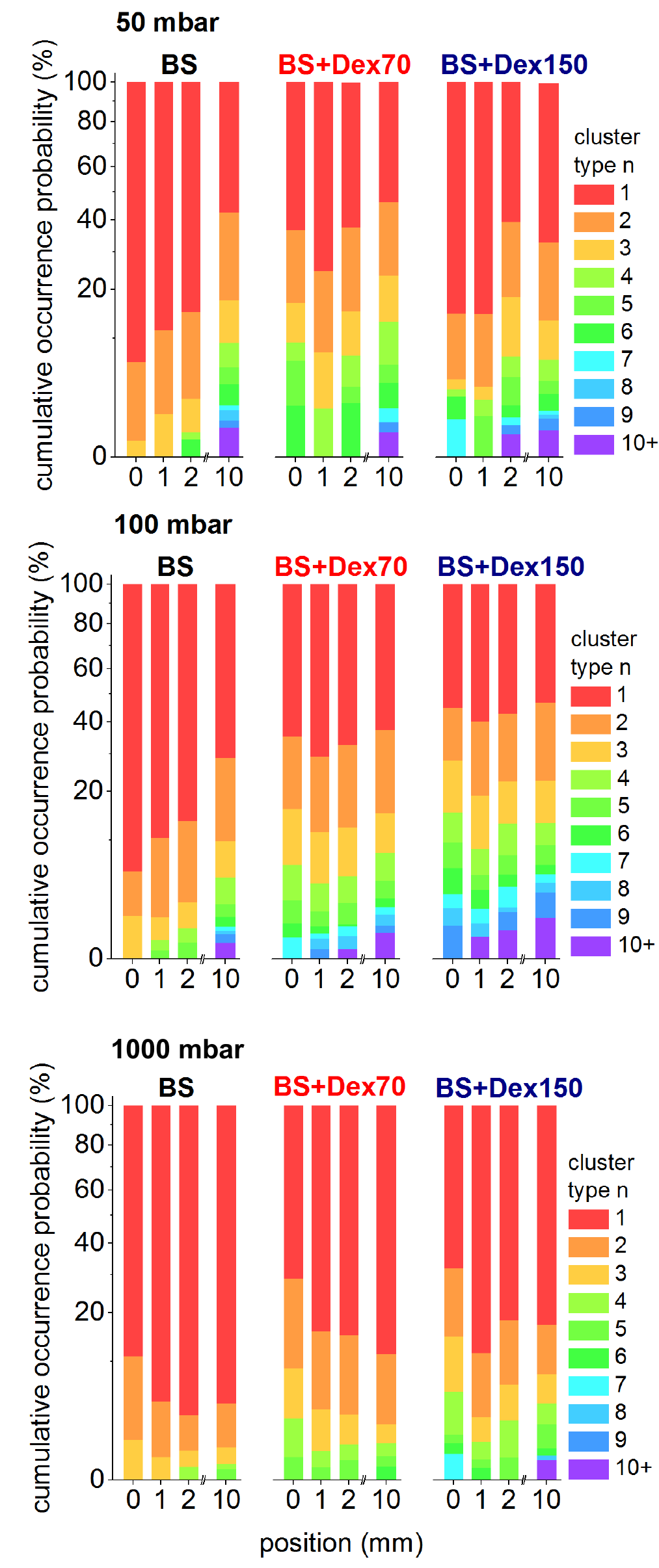}	
 	\caption{\label{fig:cluster_occ}[Color online]  Cumulative occurrence probability of different cluster types $n$ per solution at four different distances from the entrance along the channel for $\Delta P = 50$, $100$ and \unit{1000}{mbar}. Type 1 indicates single cells.}
 \end{figure}

Cluster formation in our experiment is a dynamic process and the total amount of cells of a cluster increases down the channel. To quantify this evolution, we maintained the tube hematocrit in the range of $0.4-0.8$\%, where no significant differences in cluster formation were observed, and we determined the occurrence of clusters type \textit{n} at each channel position. At $\Delta P =$ \unit{100}{mbar}, most cells enter the channels as singlets ($n = 1$). For the BS solution, a few clusters of type $n = 2$ and $3$ are observed. In the presence of macromolecules rouleaux enter the channel and clusters of type $n > 3$ are observed. The same behaviour is observed for $\Delta P = 50$ and \unit{1000} {mbar} (cp. Fig.\ref{fig:cluster_occ}). Afterwards, there is a monotonic increment in the cluster type $n$ down the channel in all cases.

This observation in the BS solution clearly indicates that hydrodynamic interactions induce cluster formation along the channel. This conclusion is also true for the solutions with dextran, where a few clusters appear to break up between positions 0 and 1 at first, but more clusters subsequently form down the channel. One may speculate from the larger number of the polymer-induced clusters that they should also be more robust against additional perturbations from irregularities in the flow, but this hypothesis should be verified in future studies.

Interestingly, while clusters build up progressively as the cells flow along the channel because of hydrodynamics interactions, increasing the flow strength does not intensify this phenomenon: to see that, one must compare configurations after equal residing time in the channel, e.g. position 1 at $\Delta P=$\unit{100} {mbar} (hereafter called the low stress configuration) and position 10 at $\Delta P=$\unit{1000} {mbar} (hereafter called the high stress configuration). Firstly, it is interesting to observe that, in the BS solution, the cluster size distributions are relatively close in both low and high stress configurations, as would be expected for objects of fixed shapes. This indicates that in the range of physiological velocities, flow-induced deformations of cells and the subsequent hydrodynamics interactions do not depend much on the flow stress. On the contrary, in the macromolecules cases, we observe a strong increase of the number of isolated cells between the low and the high stress configurations. This can be seen as a reminiscence of the well-known breaking of rouleaux by shear flow in unconfined configurations. Note that nevertheless as already highlighted in Ref. \cite{Brust2014}, rouleaux remain present even in high stress configuration, although the mean shear rate is much higher than the typical shear rate of \unit{10}{s}$^{-1}$, at which one generally considers all rouleaux broken under simple shear flow in bulk.

\subsection{Cluster length characterization.}

In order to go one step further in the quantification of the effect of flow on cluster formation, we need to describe more accurately those clusters.

\begin{figure}[h!]
	\centering
	\includegraphics[width = 0.75 \linewidth]{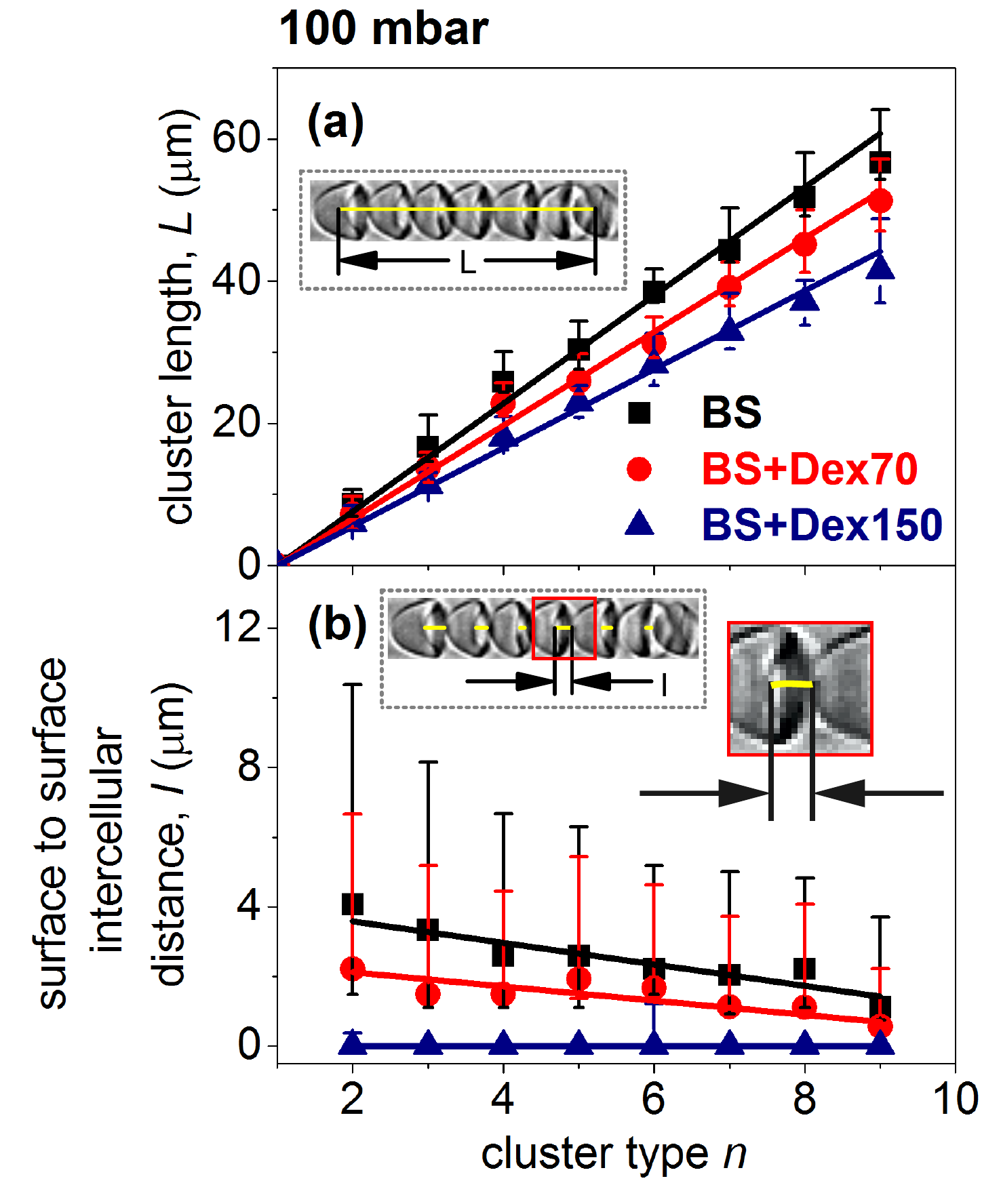}
	\caption{\label{fig:inter} [Color online]  ($\Delta P =$ \unit{100}{mbar}) \textbf{(a)} Median value $\widetilde{l}$ of the cluster length $L$ distribution by cluster type $n$. The error bars are the positions of the first and third quartil. \textbf{(b)} Median value of the surface-to-surface inter-cellular distance $I$ over the axial line per cluster type $n$. The error bars represent the positions of the first and third quartil. Straight lines indicate linear fits.}
\end{figure}

We first quantify the differences between cluster types by their lengths, which are defined as the distance between the centre of the first and last cells of the cluster (see photograph at the upper left corner in Fig.\,\ref{fig:inter} (a)). Statistics on this length parameter as well as on the other cluster characteristics that are considered in the following are obtained by considering indiscriminately positions 1, 2 and 10.

Each data point on the graph (a) in Fig.\ref{fig:inter} represents the median value $\widetilde{l}$ of the cluster length distribution by cluster type, and the error bars of their first and third quartil. If we take the case of $\Delta P =$ \unit{100}{mbar} again, we observe that the length difference between the BS and BS+Dex70 (Dex150) solutions is $15$\% ($33$\%), and between the two solutions with macromolecules is $16$\%. These length differences are significant but are a consequence of two opposite effects: the cells in the cluster in the BS solution have a parachute-like shape, but some finite inter cell distances, whereas the cells in the solutions with macromolecules are more elongated, but closely attached to one another. This latter property becomes apparent if we look at the surface-to-surface inter-cellular distance $I$ in Fig.~\ref{fig:inter}(b). The inter-cellular distances $I$ are asymmetrically distributed and we again take the median as a representative value. The 1st and 3rd quartiles of the distribution are shown as bars up and above the median value on the graph. We found that for the surface-to-surface inter-cellular distance in the BS solution, the distance starts from $I = 4.1^{+6.3}_{-2.6}\,\textrm{\micro\meter}$ for the cluster type $n = 2$ and $I = 2.2^{+4.4}_{-0.0}\,\textrm{\micro\meter}$ in BS+Dex70, which decreases in both cases for larger cluster types. In comparison, it is practically zero in BS+Dex150 with $I = 0.00^{+0.4}_{-0.0}\,\textrm{\micro\meter}$ for the cluster type $n = 2$ and remain similar for larger cluster types. This result indicates that the clusters in the BS+Dex70 solution are often a mixture of hydrodynamic and macromolecule-induced clusters. Additionally we present in Table\,\ref{tab:tablemedian} the median $\widetilde{l}$ and the first and the third quartil of the cluster length distribution per pressure drop for the cluster type $n = 2$, $3$ and $4$, which are clusters of the most frequent occurrence for the pressure drops studied. The median value  $\widetilde{l}$ of the cluster length $L$ increases with pressure drop due to the larger deformation of the cells at larger velocities.

\begin{table}[h!]
		\small
		\caption{\ \ Median value $\widetilde{l}$ of the cluster type $n = 2, 3$ and $4$ length distribution for each pressure drop per solution.}
		\label{tab:tablemedian}
		\begin{tabular*}{0.5\textwidth}{@{\extracolsep{\fill}}cccc}
			\hline
			$\Delta P$ & $\widetilde{l}_{n = 2}$ &  $\widetilde{l}_{n = 3}$ &  $\widetilde{l}_{n = 4}$ \\
			($\textrm{\milli}$bar) & ($\textrm{\micro\meter}$) & ($\textrm{\micro\meter}$) & ($\textrm{\micro\meter}$)		 \\
			\hline
			&  BS & BS  & BS \\
			\hline
			20  & $7.96^{+1.76}_{-1.48}$ & $15.00^{+3.52}_{-2.59}$ &  $21.85^{+4.07}_{-3.33}$ \\
			50  & $8.15^{+1.48}_{-1.11}$ & $15.59^{+1.39}_{-1.57}$ & $23.70^{+1.67}_{-2.69}$ \\
			100 & $8.61^{+2.07}_{-1.65}$ & $16.66^{+4.54}_{-2.04}$ & $25.78^{+4.28}_{-1.44}$  \\
			1000 & $11.11^{+3.15}_{-1.85}$ & $19.48^{+3.15}_{-3.33}$ & $28.00^{+4.63}_{-4.63}$  \\
			\hline
			& BS+Dex70 & BS+Dex70 & BS+Dex70\\
			\hline
			50  & $5.85^{+2.04}_{-2.22}$ & $10.78^{+3.52}_{-2.59}$ & $19.26^{+5.37}_{-4.82}$  \\
			100 &  $7.24^{+2.44}_{-2.14}$ &$13.73^{+2.22}_{-2.04}$ & $22.78^{+2.89}_{-1.78}$  \\
			1000 & $10.26^{+2.78}_{-1.11}$ & $18.43^{+4.63}_{-3.80}$ & $26.96^{+2.65}_{-6.05}$  \\
			\hline
			& BS+Dex150 & BS+Dex150 & BS+Dex150\\
			\hline
			50  & $4.44^{+1.86}_{-0.64}$ & $8.80^{+1.29}_{-1.30}$ & $12.41^{+2.03}_{-0.74}$  \\
			100 &  $6.03^{+2.41}_{-1.29}$ & $11.37^{+1.67}_{-1.11}$ & $18.00^{+2.89}_{-1.78}$  \\
			1000 & $9.44^{+2.41}_{-1.29}$ & $16.48^{+3.15}_{-1.66}$ & $25.19^{+5.00}_{-3.52}$  \\
		\end{tabular*}
	\end{table}
	
\subsection{Separating hydrodynamic and macromolecule induced clusters.}

\begin{figure}[t!]
	\centering
	\includegraphics[width = 1 \linewidth]{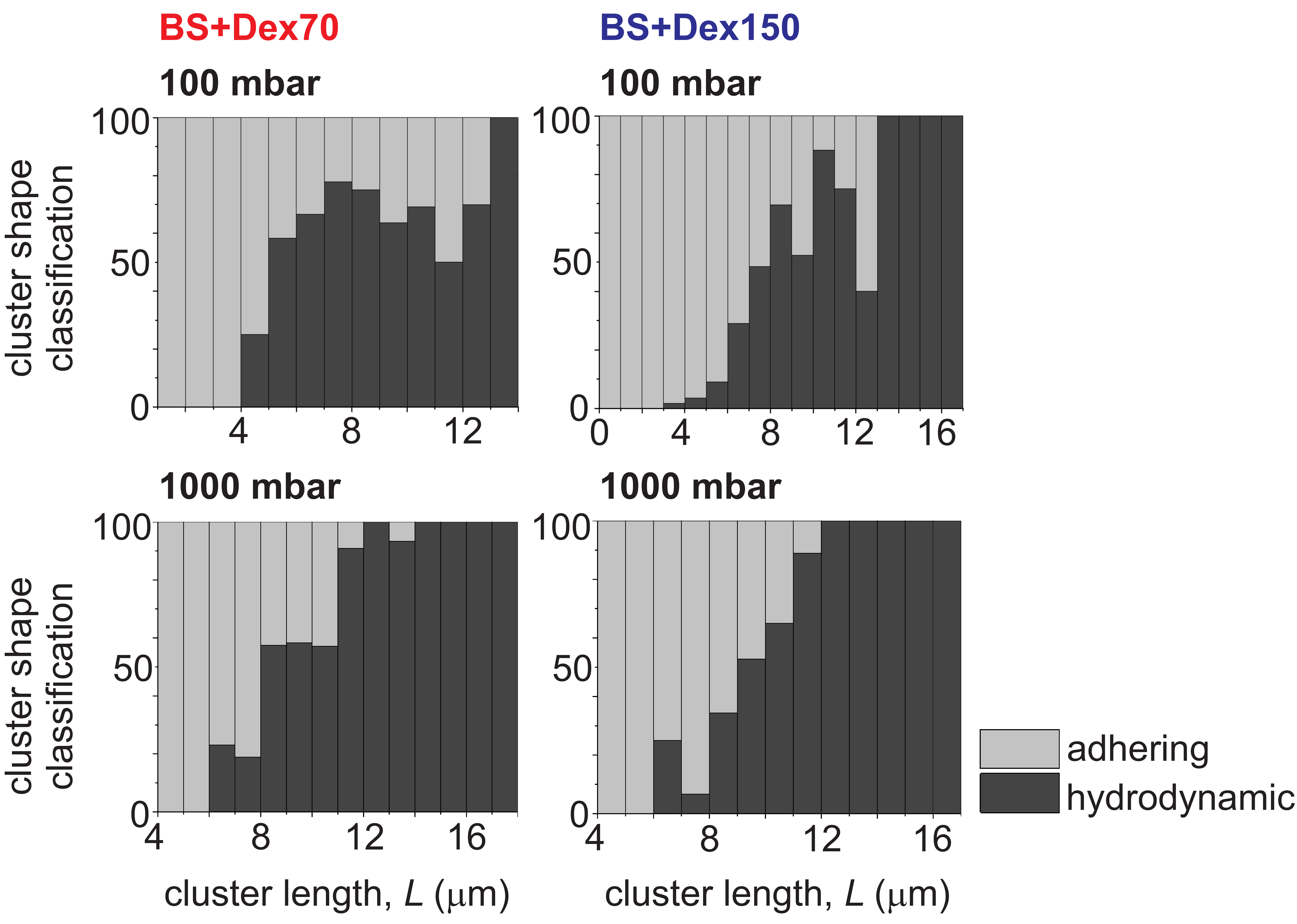}
	\caption{\label{fig:macro_shape_clas} Percentage of the hydrodynamic and adhering cluster type $n = 2$ occurrence per cluster length in BS+Dex70 (left) and BS+Dex150 (rigth) solutions flowing at a pressure drop of $\Delta P = \unit{100}{mbar}$ and $\Delta P = \unit{1000}{mbar}$.}
\end{figure}

As mentioned, clusters in macromolecule solutions are a mixture of hydrodynamic and macromolecule-induced clusters. To quantify its occurrence, we focus our analysis on the clusters type $n = 2$, as this cluster type has the highest cluster occurrence in comparison to other cluster types (cp. Fig.\ref{fig:cluster_occ}). A hydrodynamic cluster type $n = 2$ is defined as a cluster with non-zero surface to surface inter-cellular distance between the cells. More precisely, the inter-cellular distance was measured from the external border of the back cell to the inner border of the front cell (where a “rim” seems to be located) , as it is shown in the zoom of the inset in Fig.\ref{fig:inter} (b)). By a way of contrast,  macromolecule-induced or adhering clusters due to the presence of BS+Dex70 and Dex150 are known to have at rest a surface to surface inter-cellular distance of around $\unit{25}{\nano\meter}$ \cite{chien1973}. Consequently an adhering cluster in our experiments is defined as a cluster with an evidently close to zero inter-cellular distance between the cells (cp. cluster type $n = 2$ in Fig.\ref{fig:geo_clust} for BS+Dex70 and BS+Dex150 solutions). A relative occurrence of hydrodynamic and adhering clusters can now be quantified among type $n=2$ clusters. As an example, we show in Fig.\ref{fig:macro_shape_clas} the plot of the occurrence of hydrodynamic and adhering clusters for the cluster type $n = 2$ in BS+Dex70 and BS+Dex150 solutions flowing at a pressure drop  $\Delta P = \unit{100}{mbar}$ and $\unit{1000}{mbar}$, as a function of cluster length. Apparently, at larger pressure drop the relative amount of hydrodynamic formed clusters increases.

\begin{figure}[t!]
	\centering
	\includegraphics[width = 0.6 \linewidth]{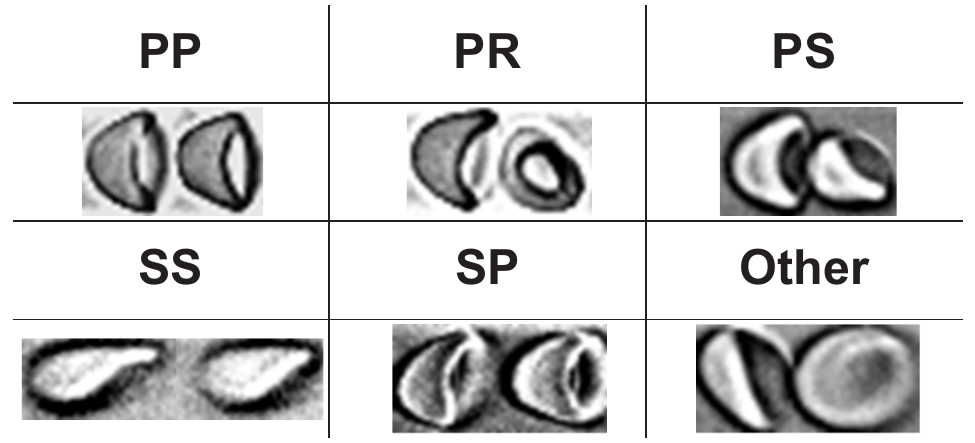}
	\caption{\label{fig:shape_clas} Cluster shape classification. PP indicates parachute-parachute, PR parachute-round, PS parachute-slipper, SS slipper-slipper and SP slipper-parachute front-and-back cell shape and Other if previous classification was not appropriate. A representative photograph per classification is presented.}
\end{figure}

\begin{figure}[t!]
	\centering
	\includegraphics[width = 0.8 \linewidth]{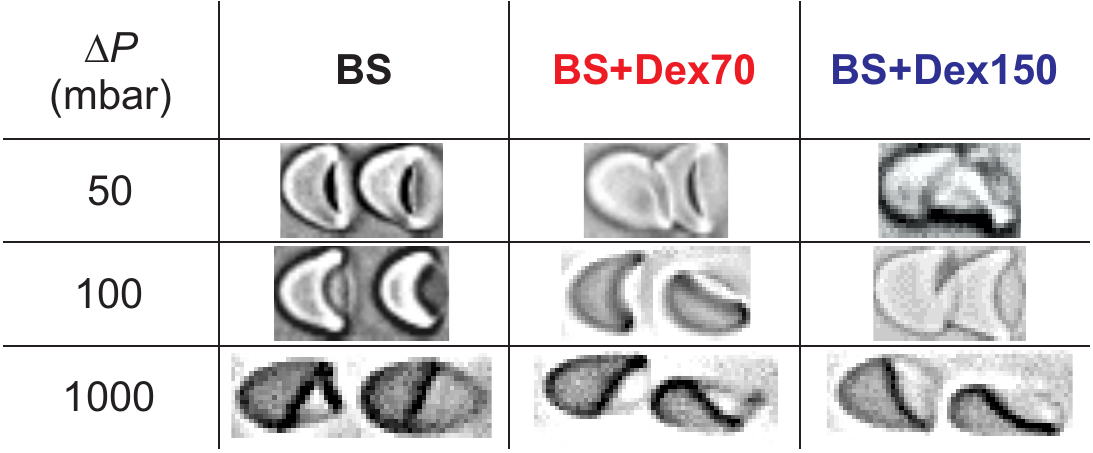}
	\caption{\label{fig:shape}  Photographs of the typical cluster configuration as a function of the pressure drop and solution for the median value  $\widetilde{l}$ of the cluster type $n = 2$ length distribution presented in table \ref{tab:tablemedian}.}
\end{figure}

We then classify the hydrodynamic cluster type $n = 2$ by visual inspection according to the shape of the front-and-back cell belonging to the cluster with parachute-parachute (PP), parachute-round (PR), parachute-slipper (PS), slipper-slipper (SS), slipper-parachute (SP) and Other, if the previous classification was not appropriate. Fig.\,\ref{fig:shape_clas} shows representative photographs of the cluster classification. Hydrodynamic clusters of type $n = 2$ were classified accordingly in each solution. From these cluster shape classifications, it is now possible to infer the typical cluster configuration for each median cluster length in Table\,\ref{tab:tablemedian} for $n = 2$. For $\Delta P = \unit{50}{mbar}$, the typical BS cluster configuration is PP, and adhering clusters for BS+Dex70 and BS+Dex150 solutions. When $\Delta P$ is increased to $\unit{100}{mbar}$, the typical BS cluster configuration remains PP, and only in BS+Dex150 solution does the cluster configuration remain adhering. In contrast, the main configuration is now PS in BS+Dex70 solution. Remarkably, at $\Delta P = \unit{1000}{mbar}$ the main cluster configuration is SS, independent of the solution. This result is summarized in Fig.\ref{fig:shape}, where a photograph of typical cell configuration per pressure drop with different solutions is presented.

These observations confirm that, while hydrodynamic interactions tend to create the so-called hydrodynamics clusters, and perhaps clusters where cells are so close that they would eventually collapse onto a macromolecules induced cluster, they do not succeed in counterbalancing the breaking of macromolecules induced clusters by flow stress: in the BS+Dex70 solution (resp. BS+Dex150) the occurrence of hydrodynamic clusters rises from 41\% to 59 \% (resp. 29\% to 55\%) when $\Delta P$ is changed from $\unit{50}{mbar}$ to $\unit{1000}{mbar}$.

\subsection{Numerical simulations of hydrodynamic and macromolecule induced clusters}

\begin{figure}
	\centering
	\includegraphics[width = 1 \linewidth]{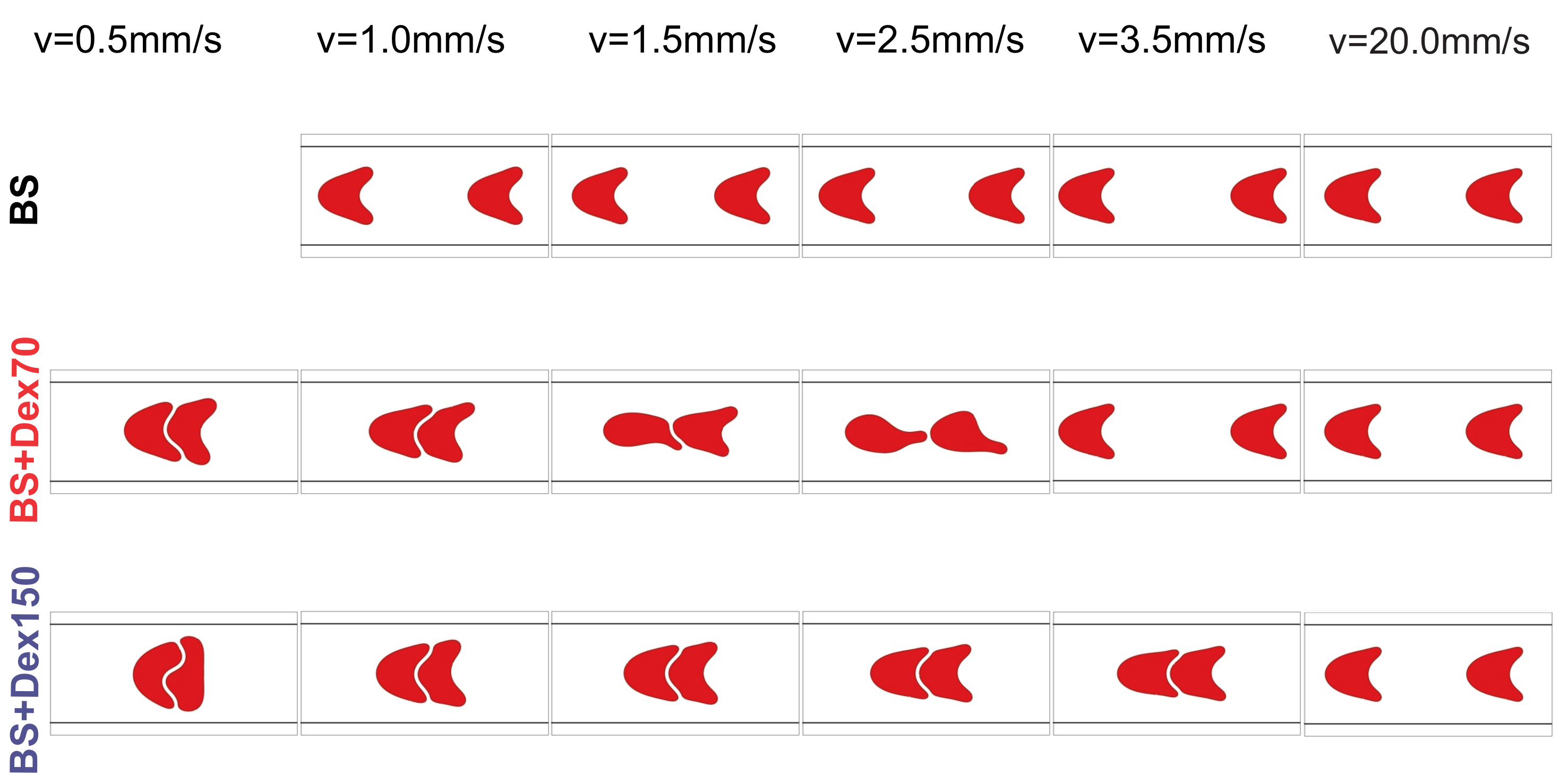}
	\caption{\label{fig:simulations} [Color online]  Numerically obtained stationary shapes of cluster type $n = 2$ at different maximum centerline flow velocities for different solutions.}
\end{figure}

To verify the effect of the hydrodynamic interaction compared to the macromolecule-induced aggregation, we perform two-dimensional numerical simulations for six different maximum centerline flow velocities of the undisturbed Poiseuille flow as described in the numerical methods section. We consider two vesicles with a radius of \unit{3}{\micro\meter} and a reduced area of $\tau = 0.65$; the vesicles are suspended in a Poiseuille flow between two parallel walls with a distance of \unit{12}{\micro\meter} at low Reynolds number. The physical parameters of the RBC are first selected such that the bending rigidity is $\kappa =$ \unit{4 \times 10^{-19} }{J} and the inner viscosity is $\eta_{i} =$ \unit{5}{mPas}. The effect of the macromolecules is mimicked by an inter-cellular interaction force derived from a Lennard-Jones potential with interaction energies $\epsilon_{Dex70} =$\unit{4.8}{\micro\joule\per\meter\squared} for the BS+Dex70 solution, and $\epsilon_{Dex150} =$\unit{12}{\micro\joule\per\meter\squared} for the BS+Dex150 solution \cite{Steffen2013}. Our simulations reproduce the effect of an additional interaction energy (adhesion energy between cells) in the cluster formation (cp. Fig.\,\ref{fig:simulations}). Notable for the case of strong adhesions (BS+Dex150 solution): the cells stick close to one another and share a significant part of their membranes in the interaction area. The interaction area depends strongly on the flow velocity, which decreases when the flow velocity increases. For the case of intermediate interaction energies (BS+Dex70 solution), the two cells stick close to one another and share part of their membranes only at low velocities (in the simulations up to $v = 2.5$ $\textrm{\milli\meter\per\second}$). At higher velocities, the cells are detached and flow separately in a hydrodynamic cluster configuration, even at flow velocities of $v = 20.0$ $\textrm{\milli\meter\per\second}$. In a pure hydrodynamic flow with zero adhesion energy between the cells (BS solution), the cells flow separately, independent of the velocity with a non-zero inter-cellular distance between the cell surfaces. Here we have to mention that in the numerical study the definition of a hydrodynamical cluster differs from the experimental results. The numerical simulation allows placing two vesicles at almost arbitrary (if not too far) distances and they always reach their shown stable inter-cellular distance. This means that cells that are placed closer to each other repel and if placed further away they attract, yielding a kind of an effective hydrodynamic interaction potential. In this way we find hydrodynamical induced clusters for all flow rates for the BS solution, but we also find that in the numerical simulations the polymer induced adhering clusters should become unstable at flow rates of 3.5 mm for the BS+Dex70 and 20 mm for the BS+Dex150 solutions. This is at least in qualitative agreement with our statistics on the occurrence of cluster types (Fig 5) where we find that at larger flow rates the percentage of hydrodynamic (non-adhering) clusters increases.
Our simulations reproduce the effect of macromolecule induced clusters in an almost quantitative manner, and we also find stable hydrodynamic clusters. However, the shape configurations at larger velocities are not well reproduced and we therefore concentrate on the experimental results for a further analysis of the hydrodynamic induced clusters.

\subsection{Probability density functions of the cluster length}
\begin{figure}[h!]
		\centering
		\includegraphics[width = 0.7 \linewidth]{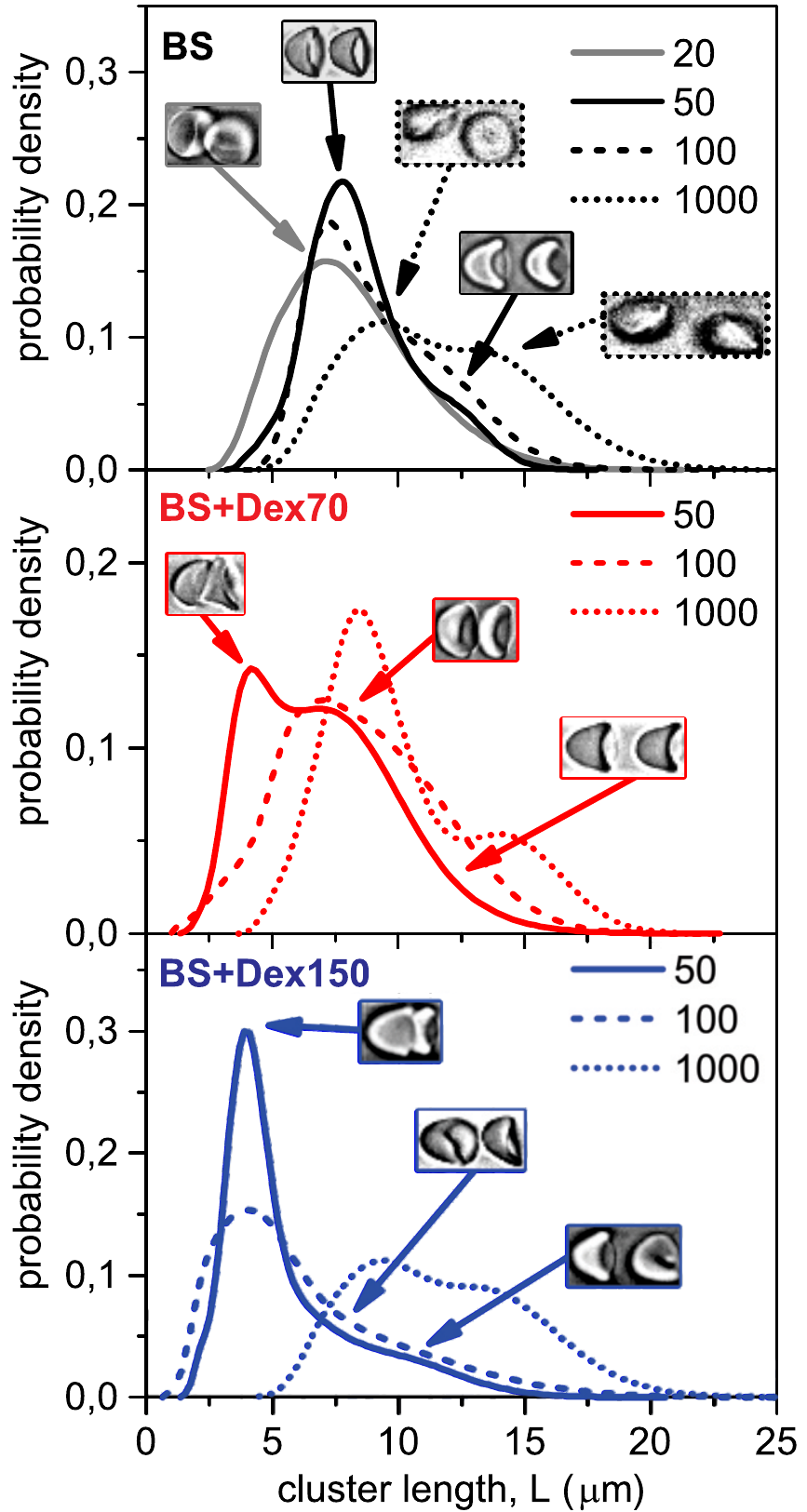}
		\caption{\label{fig:pdf_CT2} [Color online] Probability density of the characteristic cluster length $L$ of the cluster type $n = 2$ for BS, BS+Dex70 and BS+Dex150 solutions. The photographs are representative snapshots of the most frequent cluster configurations with the cluster length position indicated by the arrows between the photographs and the graph lines.}
	\end{figure}

After we have clearly separated the effect of hydrodynamic and macromolecule induced clusters, we can finally present a noteworthy result for the pure hydrodynamic clusters that was obtained by evaluating the probability density function of the cluster length $L$ of cluster type $n = 2$ (Fig.\,\ref{fig:pdf_CT2}). The results in BS solution show that, for $\Delta P = \unit{20}{mbar}$, there is a preferential position for the cluster length at $7.96^{+1.76}_{-1.48}$ $\textrm{\micro\meter}$. When we increase $\Delta P$ to $\unit{50}{mbar}$, the cell-to-cell distances in the BS solution start to be bimodal distributed: a pronounced peak for (short) distances comparable to the $\Delta P = \unit{20}{mbar}$ that is observed experimentally, but also a populated position for large distances appears in the form of a shoulder on the right-hand side of the graph (cp. Fig.\ref{fig:pdf_CT2}). This bimodal distribution is enhanced when we increase the $\Delta P$ to $\unit{100}{mbar}$, where less clusters has a short cell-to-cell distance, but more are found to have a larger distance between them. Finally, if we increase the pressure drop to $\unit{1000}{mbar}$, not only this behaviour persists, but the cluster length is also shifted to higher values. The positions of the peaks of these bimodal distributions are found by fitting a double-peak function (the sum of two bi-Gaussian functions) on the cluster length distribution. 
For the solutions with macromolecules, a third peak appears in the probability density function of the cluster length due to the adhering cells. A clear example can be seen in the cluster length distribution of the cells flowing at $\Delta P = \unit{50}{mbar}$ in BS+Dex70 solution in Fig. 9, where a photograph of a typical adhering cluster is shown, too.
	
\begin{figure}
	\includegraphics[width = 1 \linewidth]{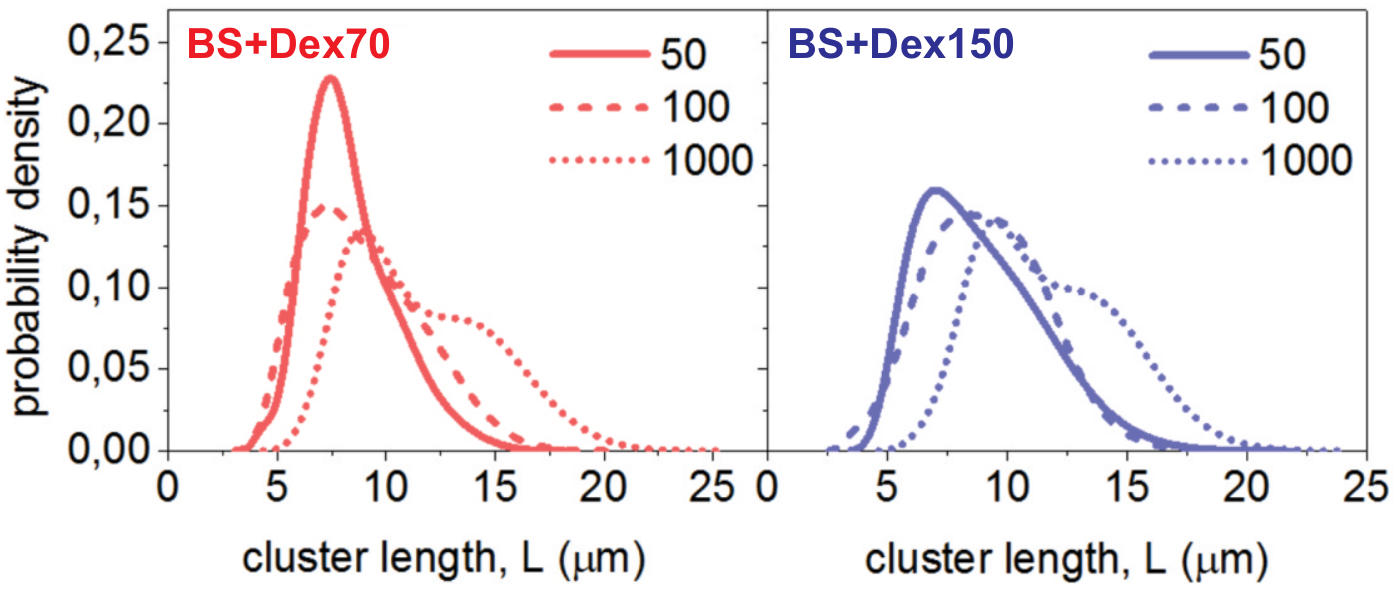}
	\caption{\label{fig:pure_hydro} [Color online] Probability density of the cluster length distributions for the pure hydrodynamic clusters type $n = 2$ in the macromolecule solutions BS+Dex70 (left) and BS+Dex150 (right). }
\end{figure}

For these solutions with macromolecules it is nevertheless also possible to determine the probability density function of the cluster length $L$ for the hydrodyanmic clusters only, as it is shown in Fig.\ref{fig:pure_hydro}. The probability density of the cluster length $L$ of the BS+Dex70 and BS+Dex150 solutions are very similar to the probability density of the pure hydrodynamic clusters formed in BS solutions (upper graph, Fig.\ref{fig:pdf_CT2}). Cell-to-cell distances in these cases are also bimodal distributed and again, a pronounced peak for short distances is observed, except in the length distribution at $\Delta P = \unit{50}{mbar}$ in BS+Dex150 solution, where a high population of clusters with a short cluster length were adhering (lower graph, Fig. \ref{fig:pdf_CT2}). This illustrates the subtle imbrication of hydrodynamics effects and short range adhesion forces, the former acting as a precursor mechanism for bringing the cells close enough for adhesion to occur, unless shear stress overcomes this process.

The positions of the peaks of the double-peak fitting function of the hydrodynamic cluster length distribution in Fig.\ref{fig:pdf_CT2}, upper graph (BS) and Fig.\ref{fig:pure_hydro} are presented in table\,\ref{tab:table3}.

\begin{table}[b!]
	\small
	\caption{\ Characteristic lengths of the hydrodynamic cluster type $n = 2$. The length distributions are fitted with a double-peak function composed of two bi-gaussian distribution of center $L_i$, and left and right standard deviation $w_i^-$ and $w_i^+$.}
	\label{tab:table3}
	\begin{tabular*}{0.5\textwidth}{@{\extracolsep{\fill}}lllllll}
		\hline
			& \multicolumn{3}{c}{peak 1} & \multicolumn{3}{c}{peak 2}\\
		$\Delta P$&$L_1$&$w_1^-$&$w_1^+$&$L_2$&$w_2^-$&$w_2^+$\\
		(mbar) &($\textrm{\micro\meter}$)&($\textrm{\micro\meter}$)&($\textrm{\micro\meter}$)&($\textrm{\micro\meter}$)&($\textrm{\micro\meter}$)&($\textrm{\micro\meter}$) \\
		\hline
		& \multicolumn{3}{c}{BS} & \multicolumn{3}{c}{BS} \\
		\hline
		20 & 6.61&1.64&3.47 & -&-&-  \\	
		50 & $7.67$&{1.37}&1.52 & $11.56$&1.61&1.64\\
		100 & $6.98$&0.93&1.48&10.43&1.75&2.63\\
		1000 & $8.4$&1.47&2.04 & $13.44$&2.94&2.92\\ 	
		\hline
		&  \multicolumn{3}{c}{BS+Dex70} & \multicolumn{3}{c}{BS+Dex70}\\
		\hline
		50  & $7.00$&0.94&1.72& $10.69$&3.06&1.70\\
		100  & $6.72$&1.33&1.68 & $10.14$&1.62&2.77\\
		1000 & $8.56$&1.27&1.76&$13.51$&2.59&2.96\\ 	
		\hline
		&  \multicolumn{3}{c}{BS+Dex150} & \multicolumn{3}{c}{BS+Dex150}\\
		\hline	
        50  & $6.52$&1.03&2.92 &$11.26$&2.14&2.41\\
		100 & $7.30$&1.62&1.66&$10.46$&1.59&1.96\\
		1000 & $9.46$&1.49&1.80& $13.54$&1.41&2.55\\
		\hline		
	\end{tabular*}
\end{table}

\begin{figure}
	\includegraphics[width = 1 \linewidth]{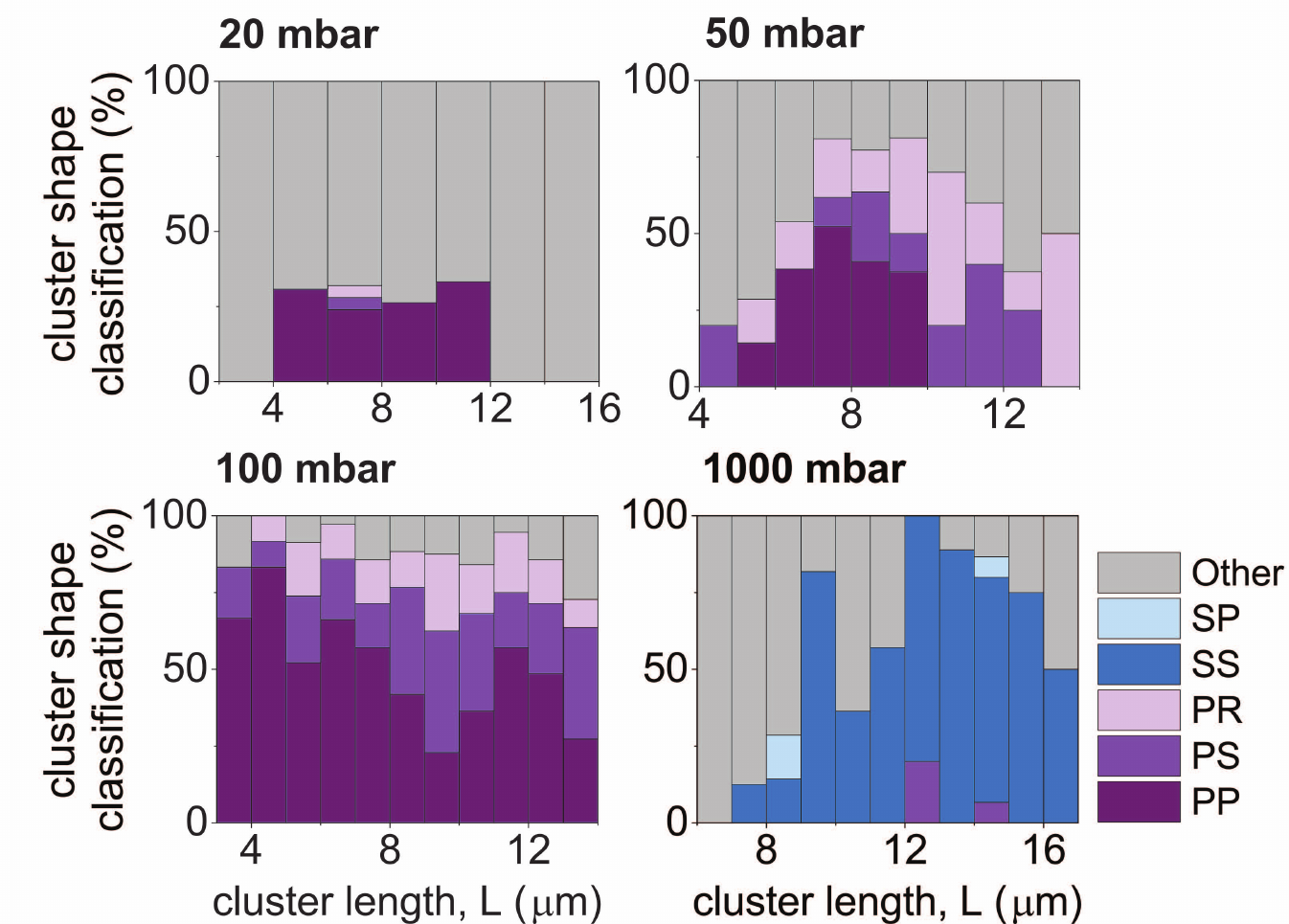}
	\caption{\label{fig:pure_hydro_clas_shape} [Color online] Percentage of the different shapes of clusters of type $n = 2$ n(cp. Fig.\ref{fig:shape}) per cluster length (BS solution).}
\end{figure}

The hydrodynamic clusters of the cluster type $n = 2$ in BS solution were classified according to their length, following the shape classification of Fig.\ref{fig:shape_clas}. The resulting cluster-shape classification are presented in Fig.\ref{fig:pure_hydro_clas_shape}. For lower pressure drop ($\Delta P = \unit{20}{mbar}$), the cells are mainly in a rather undefined configuration and are classified as Other. However, in the peak zone (cluster length $L = 6.61^{+3.47}_{-1.64}$ $\textrm{\micro\meter}$, cp. Fig.\ref{fig:shape_clas}), PP clusters are observed. When we increase the pressure drop to $\unit{50}{mbar}$, the peak 1 (cluster length $L = 7.67^{+1.52}_{-1.37}$ $\textrm{\micro\meter}$) of the bimodal distribution is now highly populated by PP clusters (52\% of the cluster configurations) and peak 2 (cluster length $L = 11.56^{+1.64}_{-1.61}$ $\textrm{\micro\meter}$) is populated mainly by PS configurations (40\%). At $\Delta P = \unit{100}{mbar}$ an even bigger population of PP (66\% of the clusters) are present for peak 1 (cluster length $L = 6.98^{+1.48}_{-0.93}$ $\textrm{\micro\meter}$), while for peak 2 (cluster length $L = 10.43^{+2.63}_{-1.75}$ $\textrm{\micro\meter}$) PP (36\%) and PS (32\%) have a similar frequency of occurrence. For the largest pressure drop ($\Delta P = \unit{1000}{mbar}$) the occurrences are significantly different. Here, the preferred configuration is mostly SS with a 89\% occurrence around the peak 2 (cluster length $L = 13.44^{+2.92}_{-2.94}$ $\textrm{\micro\meter}$), but only 14\% around peak 1 (cluster length $L = 8.4^{+2.04}_{-1.47}$ $\textrm{\micro\meter}$), and 71\% are classified as Other. Evidently, the PP configuration is related to smaller velocities and axial cell migration (cp. Fig.\ref{fig:dist}), while the SS configuration appears at larger velocities and is related to the cell location between the channel axis and the channel wall.

A comparable bimodal distribution for hydrodynamic clusters in a confined flow has been observed in a three-dimensional numerical simulation in Ref. \cite{mcwhirter2009flow}. It was shown that the cell-to-cell distances strongly depended on the RBC shapes, and even a dynamic transition was reported between the two different cell-to-cell distances because of the flow coupling with the cell shapes. Our experiments do not resolve the temporal dynamic of the cluster formation process, but it is known that the cell shapes also depend on their physical parameters.

\section{Conclusions}
Our study shows that the flow of soft deformable objects that interact either via hydrodynamic or via polymer induced forces leads to a rich clustering scenario. It is the pronounced flow-structure or rather the flow-shape coupling that is responsible for this rich dynamic. In hard sphere suspension or emulsions, polymer induced interactions (either based on bridging or on depletion) are only in the order of 3 to 20 $k_bT$ and would never be sufficient for clusters to resist the flow forces. RBCs can deform and thus form large interaction surfaces which results in much larger interaction energies in the order of 10.000 $k_BT$. Furthermore, hydrodynamic forces in colloid or droplet systems are much more simple because they remain at least roughly spherical and the flow field is easier to determine.

In our system, we find that cluster formation in micro-capillaries under healthy physiological conditions should be caused by a combination of hydrodynamic and macromolecule-induced interactions (non-zero and close to zero inter-cellular distance between the surface of the cells, respectively). Macromolecule-induced interactions are not fully overcome by shear stresses within the physiological range, and they contribute to cluster stability. Yet, as confirmed by our two-dimensional numerical simulations, cluster stabilization by hydrodynamics might become predominant in the upper part of the physiological range of flow velocities.

The physical origin of the pronounced bimodal distribution of the cell-to-cell distances in the hydrodynamic clusters results from changes in the cell shapes. Whether the changes in shape are dynamical because of the shape-flow coupling, or because of differences in the physical properties of the cell, remains an open question. To answer this, experiments will be necessary to resolve the temporal dynamics of the cluster formation process in direct comparison with three-dimensional simulations. Furthermore, it might be worthwhile to do experiments with RBC`s that have been separated by density. This yields a separation by age and eventually leads to a more monodisperes distribution oh their physical parameters such as the bending rigidity. 

We acknowledge the support of the German Science Foundation research initiative SFB1027 and of the Region Languedoc-Roussillon and the labex NUMEV (convention ANR-10-LABX-20). Also we would like to thank Daniel Flormann from AG Wagner group of the University of Saarland, Germany, for the viscosity data for BS+Dex70 and BS+Dex150 solutions.

\footnotesize{
\bibliography{rsc} 
\bibliographystyle{rsc} 
}

\end{document}